# Consequences of the background in piezoresponse force microscopy on the imaging of ferroelectric domain structures

T. JUNGK*, Á. HOFFMANN & E. SOERGEL

Institute of Physics, University of Bonn, Wegelerstraße 8, 53115 Bonn, Germany
*corresponding author: Tel.: +49 228 73-4895 Fax.: +49 228 73-4038 E-mail-address: jungk@uni-bonn.de

**Summary**

The interpretation of ferroelectric domain images obtained with a piezoresponse force microscope (PFM) is discussed. The influence of an inherent experimental background on the domain contrast in PFM images (enhancement, nulling, inversion) as well as on the shape and the location of the domain boundaries are described. We present experimental results to evidence our analysis of the influence of the background on the domain contrast in PFM images.

**Introduction**

Ferroelectric domain patterns are the basis of a multitude of applications such as quasi–phase–matched frequency converters (Fejer et al., 1992), electro-optic scanners (Gahagan et al., 2001), nonlinear photonic crystals (Broderick et al., 2000), and ultra-high density data storage devices (Cho et al., 2005). For the visualization of ferroelectric domains several techniques have been developed (Soergel, 2005), however, domain selective etching (Barry et al., 1999) and piezoresponse force microscopy (Alexe & Gruverman, 2004) are by far the most utilized. Selective etching is popular because it gives a simple and rapid estimate of the domain structure over large areas, though it is destructive. Piezoresponse force microscopy, even though the image size is restricted to about $100 \times 100$ µm$^2$, is widely used because of its high lateral resolution and non-destructive imaging capability. Furthermore, the possibility to modify the domain structure with the help of its sharp tip (Cho et al., 2005) makes the piezoresponse force microscope (PFM) a versatile tool for the investigation of ferroelectric domains and domain boundaries. Although domain structures are easily imaged with this method, the interpretation of the obtained images, however, is still challenging because of the complexity of the detection mechanism. That is why a lot of surprising features concerning the domain contrast and the shape of domain boundaries were published (Agronin et al., 2005; Harnagea et al., 2003; Hong et al., 2002; Kolosov et al., 1995; Labardi et al., 2000; Labardi et al., 2001, Scrymgeour & Gopalan, 2005; Shvebelman et al., 2002, Xu et al., 2004).

Piezoresponse force microscopy is based on the deformation of the sample due to the converse piezoelectric effect. The PFM is a scanning force microscope (SFM) operated in contact mode with an additional alternating voltage applied to the tip. In piezoelectric samples this voltage causes thickness changes and therefore vibrations of the surface, which lead to oscillations of the cantilever that can be read out with a lock-in amplifier. The different orientations of the polar axis of adjacent domains lead to a domain contrast in PFM measurements, i.e., the domains are for example displayed as bright and dark areas in PFM images. However, in a previous paper (Jungk et al., 2006) we have shown that PFM measurements are usually governed by a frequency-dependent background which is inherent to the experimental setup. Typically the background has an amplitude in the order of 10 pm/V which is comparable to the piezoresponse of various materials such as GASH, KTiOPO$_4$, LiNbO$_3$, LiTaO$_3$ or TGS (Landolt-Börnstein, 1981). For materials with a small piezoelectric coefficient this background leads to a domain contrast whose amplitude and phase depend on the frequency of the alternating voltage applied to the tip: the measured oscillation amplitudes of the cantilever are usually larger than the theoretically expected values and the required phase shift of 180° between adjacent domains is not always obtained. Moreover, this background can influence the shape and location of domain boundaries in PFM imaging.

The aim of this contribution is to point out possible causes for the misinterpretation of PFM images and to propose experimental settings for an unambiguous data acquisition. A plausible model allows the quantitative estimate of the contribution of the background to the domain contrast. The considerations presented in this paper might provide a deeper insight into PFM measurements and should be taken into account when drawing conclusions from images of ferroelectric domains and domain boundaries obtained with piezoresponse force microscopy.

**Experimental conditions**

To utilize a SFM for piezoresponse force microscopy requires mainly two instrumental features: (i) an electrical connection to the tip and (ii) direct access to the signals of the position sensitive detector recording the movement of the cantilever. We will restrict ourselves here to the vertical cantilever movement only, i.e., its bending normal to the surface. Furthermore, a lock-in amplifier is necessary for sensitive readout of the cantilever movement.

In the following the crucial parts of the experimental setup are described in order to define the parameters and denotations used further on. In addition, our experimental settings for PFM operation are given:

- *Tip of the SFM*: For PFM operation the tip must be conductive and electrically connected to allow the application of voltages. The resonance frequency of the cantilever is not crucial; it should always be far away from the frequency of the alternating voltage applied to the tip. Typically cantilevers with resonance frequencies $f_0 > 100$ kHz are utilized. In addition, for contact mode, this resonance is noticeably shifted to higher frequencies (Rabe et al., 1996). The alternating voltage is usually chosen to have a frequency between 10 kHz and 100 kHz with an amplitude $U \leq 20$ V$_{pp}$. The time constant of the feedback-loop of the SFM must be large compared to the period of modulation of the applied voltage to avoid a compensation of the signal.

  We utilize Ti-Pt coated tips (MicroMasch) with resonance frequencies $f_0 = 150 - 400$ kHz, spring constants $k = 3 - 70$ N/m and apply an alternating voltage of $\omega \approx 38$ kHz with an amplitude of 10 V$_{pp}$.

- *Sample*: In large part PFM measurements are performed with crystals exhibiting antiparallel domains only. For investigation the samples are cut in such a way that the domain boundaries are perpendicular to the surface to be studied. We will restrict ourselves to such a configuration exclusively.

  In the experiments presented here, we used a 0.5 mm thick, $z$-cut, periodically-poled LiNbO$_3$ (PPLN) crystal with a period length of 30 μm, thus exhibiting ±$z$ domain faces.

- *SFM*: Generally all scanning force microscopes are suited for PFM operation as long as they allow application of voltages to the tip and separate readout of the cantilever movement. The scanning velocity has to be adapted to the rise time of the lock-in amplifier.

  We use a SMENA SFM (NT-MDT), modified to apply voltages to the tip and upgraded with an additional interface board for readout of the cantilever movement. Typical scanning velocity is about 1 μm/s.

- *Lock-in amplifier*: Most PFM setups use dual–phase lock-in amplifiers which allow to chose between two output schemes: (i) in–phase output (also denoted as $X$-output) and orthogonal output ($Y$-output) or (ii) magnitude $R = \sqrt{X^2 + Y^2}$ and phase $\theta = \arctan(Y/X)$. These output signals of the lock-in amplifier will be named PFM signals: **P** on a positive +$z$ domain face and **N** on a negative –$z$ domain face. To specify the output (and thus the component of the particular vector) the adequate symbol ($X$, $Y$, $R$ or $\theta$) will be added as a subscript. For example $P_X$ denotes the in-phase output signal of the lock-in amplifier on a positive +$z$ domain face.

  The experiments presented in this contribution are performed with a SR830 lock-in amplifier (Stanford Research Systems). Typical settings are 1 mV for the sensitivity and 1 ms for the time constant.

The aim of PFM measurements is to detect a deformation of the sample due to the converse piezoelectric effect. The response, i.e. the thickness change of the crystal, will be denoted as the piezoresponse signal **d**. Depending on the orientation of the polar axis, **d** is either in phase or out of phase by 180° with respect to the alternating voltage applied to the tip. Unfortunately, PFM measurements are generally dominated by a system-inherent, frequency-dependent background (Jungk et al., 2006). The background signal, in the following denoted as **B**, can be expressed as the average of the PFM signals on a +z and a –z domain face: **B** = ½ (**P** + **N**). Note that both, the piezoresponse and the background signal, are not directly accessible, but they have to be calculated from the measured PFM signals.

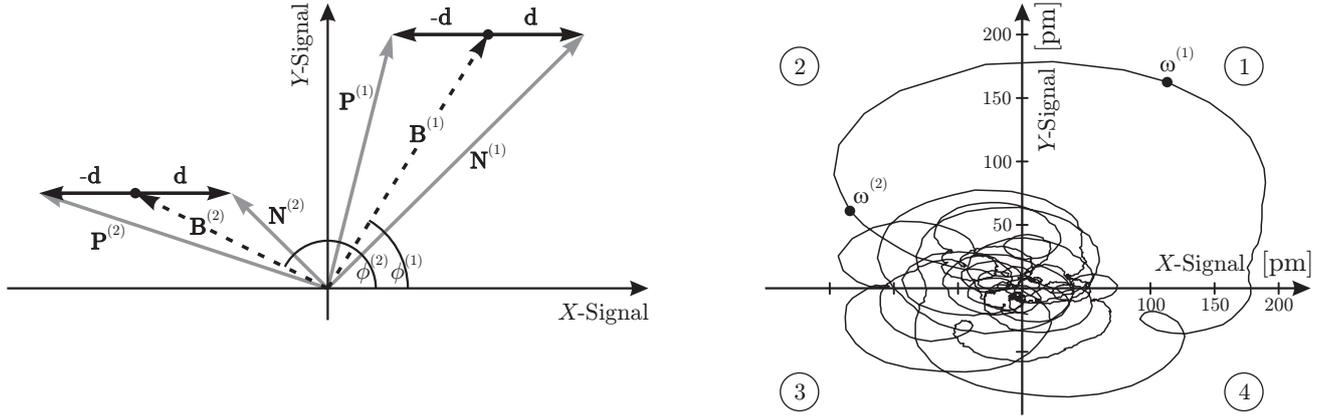

**Fig. 1.** (a) Vector diagram showing the different PFM signals on the ±z domain faces of a ferroelectric sample (X and Y: in–phase and orthogonal output of the lock-in amplifier) for two different frequencies $\omega^{(1)}$ and $\omega^{(2)}$. The superscripts indicate the corresponding signals: The PFM signal **P** (**N**) measured on a +z (–z) domain, the PFM background **B** and its appropriate phase $\phi$. The piezoresponse signal from a ±z domain face is denoted by $\mp$ **d**. (b) Frequency dependence (10 – 100 kHz) of the background **B**, determined on a PPLN surface with 10 $V_{pp}$ applied to the tip. The circled numbers 1 to 4 indicate the four quadrants.

**Vectorial description of the PFM detection**

For the correct interpretation of PFM measurements it is necessary to take into account the full data content of the PFM signals, i.e. both, the in-phase and the orthogonal signal or the magnitude and the phase.

We therefore describe the PFM signals as vectors in the X-Y-plane of the lock-in amplifier. Figure 1(a) shows the vector diagram of PFM detection for a sample with ±z domain faces for two frequencies $\omega^{(1)}$ and $\omega^{(2)}$. The piezoresponse signals ±**d** have a phase of either 0° or 180° with respect to the alternating voltage U applied to the tip and sit on top of the system-inherent background signal **B**. The alternating voltage U defines the reference phase and thus the X-axis of the X-Y-coordinate system. Reading out the magnitude R of the lock-in amplifier leads to PFM signals $P_R = |\mathbf{P}|$ on a +z domain face and $N_R = |\mathbf{N}|$ on a –z domain face, respectively. As it can easily be seen (Fig. 1a), the magnitudes of the PFM signals $P_R$ and $N_R$ are not equal and they are both larger than the expected value d of the piezo-response signal. Moreover, their relative phase is by far not 180°, although, +d and –d exhibit a 180° phase difference. These phenomena are due to the background signal **B** that can reach amplitudes comparable to the piezoresponse signal. Note that **B** can also be separated into an in-phase component $B_X$ and an orthogonal component $B_Y$.

To illustrate the importance of the system-inherent background B, its dependence on the frequency of the alternating voltage applied to the tip is shown for one specific cantilever in Fig. 1(b). The frequency of the applied voltage is scanned from 10 kHz to 100 kHz. It is obvious that phase and amplitude of the background signal vary almost arbitrarily with frequency; **B** is distributed over all four quadrants of the coordinate plane. This background strongly depends on the frequency: the big loop has a frequency span of 3 kHz only [39 kHz – 42 kHz; see also Fig. 1(d) from (Jungk et al., 2006) which was obtained with the same cantilever]. The amplitude of the background signal scales linearly with the applied voltage and is generally ≤ 10 pm/V (except for the big loop). This is of the same order of magnitude as the piezoelectric coefficient of several ferroelectric crystals such such as LiNbO$_3$ ($d_{33} \approx$ 8 pm/V; Jazbinšek & Zgonic, 2002).

**Consequences of the background signal on PFM images**

The presence of the background signal has serious consequences on PFM measurements. Note that a little shift of the frequency of the alternating voltage applied to the tip can result in drastic changes of the background signal which in turn are followed by significant changes in the PFM images obtained with the generally used *R*-output. Several surprising features concerning the domain contrast as well as the shape and location of domain boundaries turn out to possibly originate from the system-inherent background. Of course, also physical effects can influence the domain contrast or the domain boundaries, however, a careful analysis of the measured data is mandatory to avoid misinterpretation. In the following, we exemplify some possible consequences of the background:

- Enhancement of the domain contrast
- Nulling of the domain contrast
- Inversion of the domain contrast
- Arbitrary phase difference between ±$z$ domains
- Shift of the domain boundary
- Change of the shape of the domain boundary

The domain contrast *D*, as it is observed in PFM measurements when using the magnitude output *R* from the lock-in amplifier for image acquisition, is given by $D = (N_R - P_R)/(N_R + P_R)$. From Fig. 1(a) it is obvious that *D* reaches maximum when $\phi = 0°$ and $|\mathbf{B}| \geq |\mathbf{d}|$.

A minimum of *D*, so called "nulling" of the domain contrast, is observed when $\phi = 90°$ or $\phi = 270°$. In this case $P_R = N_R$ and therefore $D = 0$; the PFM images only exhibit dark lines at the domain boundaries, where the mechanical deformation is suppressed by clamping because of the different orientation of the domains.

An inversion of the domain contrast can be observed when the background changes its sign. This is the case for example when **B** switches from quadrant 1 → 2, therefore [$N_R > P_R$] is replaced by [$N_R < P_R$] and *D* changes its sign. Thus, an unambiguous identification of ±$z$ domains becomes impossible.

Detailed considerations based on the vector diagram of Fig. 1(a) allow to understand the influence of the background signal on the domain contrast. The consequences of the background signal on the shape and location of the domain boundaries in PFM measurements when using the magnitude output of the lock-in amplifier, however, need a more careful analysis. For a better understanding, both cases the in-phase background $B_X$ and the orthogonal background $B_Y$ will be treated separately. From Fig. 1(b) it is evident that usually a mixed background is present.

For modelling, we approximate the PFM signal across a domain boundary with a hyperbolic tangent, i.e., $X = \tanh(s)$ with *s* denoting the lateral position at the sample surface perpendicular to the domain wall being located at $s = 0$, hence, the amplitude of the piezoresponse signal **d** is normalized to 1. In the following the background signals will also be given by normalized values such that a background of 1 has the same amplitude as **d**.

The width of the domain boundary, i.e., the slope of the hyperbolic tangent, is determined mainly by the tip radius (Jungk et al., 2007). It must not to be confused with the real width of the domain wall over which the polarization reverses which is known to be ≤ 2 nm (Zhang et al., 1992). Here we use a 25% − 75% criterion to determine the width of the domain boundary seen with PFM for the in-phase signal *X* that corresponds to the full width at half maximum of *R*-signal if no background is present.

**In-phase background signal**

Adding the background $B_X$ to the PFM signal leads to:
$$X = \tanh s + B_X$$
$$Y = 0 \quad \Rightarrow \quad R = |\tanh s + B_X|,$$

because $R = \sqrt{X^2 + Y^2}$. The consequences can be seen in Fig. 2(a) where scan lines across a domain boundary for both the *X*- and the *R*-output are simulated. In the case of no background signal ($B = 0$, thick lines) both readout signals show the domain boundary at its real position $s = 0$, in the $R$-signal as a minimum and in the $X$-signal as the inflection point of the slope. When adding the background signal $B_X$, the minimum of $R$ is shifted by $\Delta s$ pretending the domain boundary to be at a different location. Moreover, a distinct change of the domain contrast can be observed.

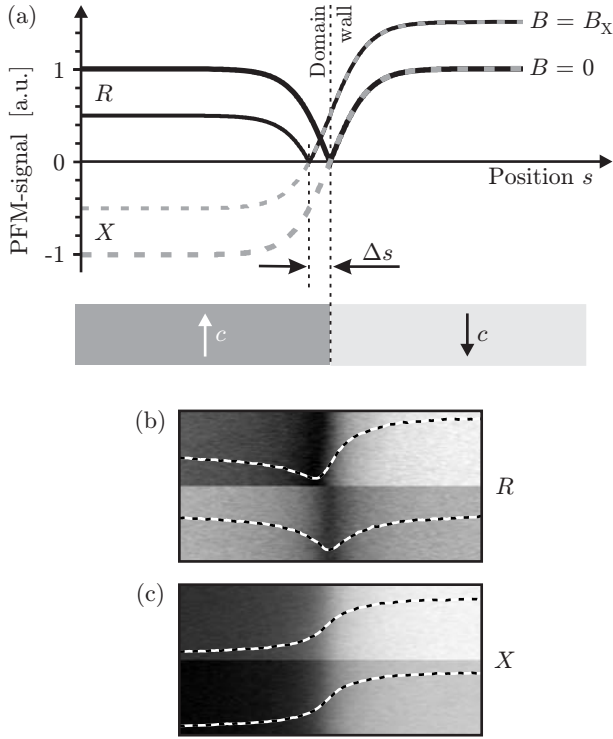

**Fig. 2.** Influence of the read-out settings of the lock-in amplifier on the detected domain wall. (a) Model predictions of the expected PFM signals where black lines correspond to the $R$-signal and grey dashed lines to the $X$-signal without ($B = 0$) and with the presence of a background signal ($B = B_X$). The model is confirmed by PFM measurements of a single domain boundary in $LiNbO_3$. During image acquisition, the frequency of the applied voltage was changed, thereby adding a background signal $B_X$. Using the $R$-output leads to a pretended shift of the domain boundary (b) whereas the PFM image recorded with the $X$-output just becomes brighter (c). The line scans are averages over 20 image lines. The image size is $1 \times 0.5$ μm$^2$.

Figures 2(b-c) show images of a single domain boundary recorded simultaneously with the $X$- and $R$-output of the lock-in amplifier. After the first half of the image, the frequency of the alternating voltage is changed in order to alter the background. Whereas in the $X$-signal the location of the domain boundary is not affected (Fig. 2c) the image taken with the $R$-output shows a distinct shift of the domain boundary (Fig. 2b). This pretended shift can be easily calculated to be

$$\Delta s = -\mathrm{artanh}\, B_X\ .$$

As a further consequence of the in-phase background a broadening and also an asymmetry of the detected domain wall is pretended. In the extreme case, when the background signal $B_X$ is larger than the piezoresponse signal $d$ no minimum can be observed in the $R$-output of the lock-in amplifier at the domain boundary. For $B_X < d$ the asymmetric broadening can be calculated by taking the full width at half maximum for each side next to the minimum of the $R$-signal separately, which yields a width:

$$W(B_X) = \mathrm{artanh}\left(\frac{4}{5 - B_X^2}\right)\ .$$

To give an example, the domain wall is asymmetrically broadened by 29 nm for a background signal $B_X = 0.5$ if the original width of the domain wall seen by PFM is 100 nm. For these values of the background signal and the initial wall width the domain boundary is seemingly shifted by 55 nm. Note that the apparent shift of the domain boundary depends on the tip radius, as the slope of the $X$-signal is steeper for sharper tips.

### Orthogonal background signal

Adding the background $B_Y$ to the PFM signal leads to:

$$X = \tanh s \quad\Rightarrow\quad R = \sqrt{(\tanh s)^2 + B_Y^2}\ .$$
$$Y = B_Y$$

From Fig. 3(a) it is obvious that for a background signal $B_Y$ the PFM images recorded with the $R$-output show the domain boundaries only. Their full width at half maximum $W$ can be calculated through the following formula:

$$W(B_Y) = 2\,\mathrm{artanh}\sqrt{\tfrac{1}{4} - \tfrac{1}{2}B_Y^2 + \tfrac{1}{2}\sqrt{B_Y^2 + B_Y^4}}$$

Furthermore, the contrast $C$ of the domain boundary decreases with increasing background $B_Y$. With the common definition $C$ can be calculated to be:

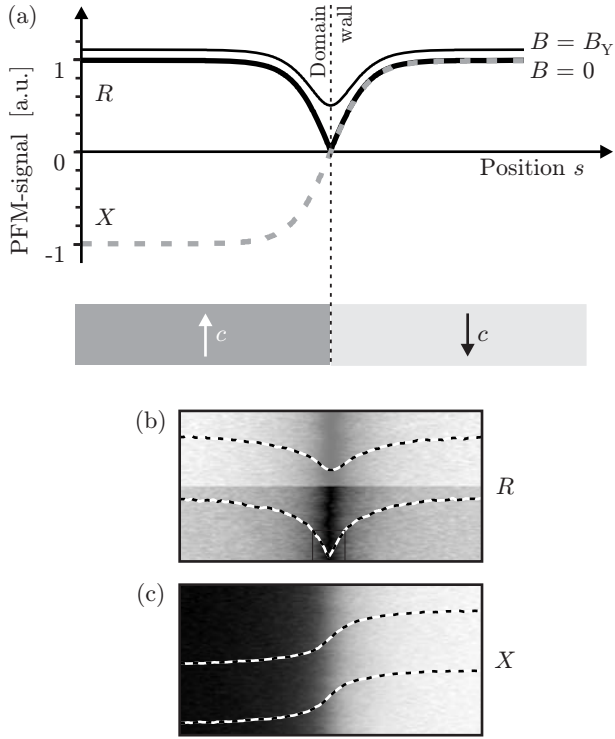

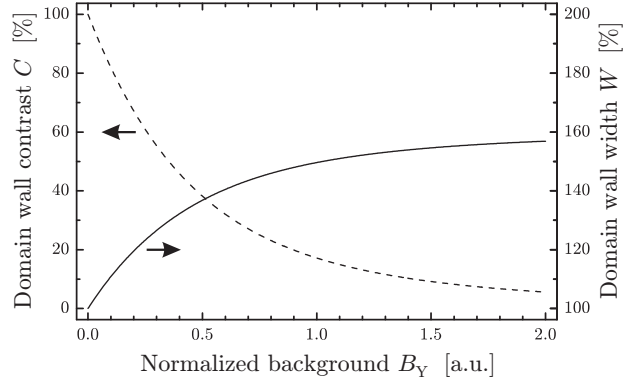

**Fig. 4.** Calculated domain wall width W (——) and domain wall contrast C (– – –) for an orthogonal background $B_Y$. The graphs are normalized to the values of the original domain wall.

**Fig. 3.** Influence of the read-out settings of the lock-in amplifier on the detected domain boundary. (a) Model predictions of the expected PFM signals where black lines correspond to the $R$-signal and grey dashed lines to the $X$-signal. The model is confirmed by PFM measurements of a single domain boundary in LiNbO$_3$. During image acquisition, the frequency of the voltage applied to the tip was changed, thereby adding a background signal $B_Y$. Using the $R$-output leads to a pretended broadening of the domain boundary (b) whereas the image of the $X$-output stays unchanged (c). The line scans are averages over 20 image lines. The image size is 1 × 0.5 µm$^2$.

$$C = \frac{R_{max} - R_{min}}{R_{max} + R_{min}} = \frac{\sqrt{1 + B_Y^2} - |B_Y|}{\sqrt{1 + B_Y^2} + |B_Y|}$$

Figures 3(b-c) show PFM images of a single domain boundary recorded simultaneously with the $X$- and $R$-output of the lock-in amplifier. After recording half of the image, the frequency of the alternating voltage is changed in order to alter the background. Whereas in the $X$-signal no changes can be observed (Fig. 3c) the image taken with the $R$-output shows a distinct broadening of the domain wall as well as a faded contrast (Fig. 3b). To give an example, the width of the domain boundary is symmetrically broadened by 37 nm for a background signal $B_Y = 0.5$ if the original width of the domain wall seen by PFM is 100 nm. Simultaneously the contrast drops down to 38%

of its initial value. Both effects, the increasing domain wall width and the decreasing domain wall contrast are shown for an orthogonal background $B_Y \leq 2$ in Fig. 4.

**Full background signal**

As it was already mentioned, usually a mixed background $B_R = \sqrt{B_X^2 + B_Y^2}$ is present in PFM measurements as it can be clearly seen from the frequency spectrum in Fig. 1(b). In this general case the $R$-signal is given by:

$$R = \sqrt{(\tanh s + B_X)^2 + B_Y^2} \; .$$

This, however, leads to a superposition of the effects described above, and can therefore have serious consequences on the pretended domain contrast as well as on the features of the domain boundary when using the $R$-output of the lock-in amplifier. The width of the domain wall for a full background **B** can be calculated as the sum of $W(B_X)$ and $W(B_Y)$ and is plotted in Fig. 5(a).
With the full background also the phase shift $\Delta\theta$ between $+z$ and $-z$ domain faces can be calculated via:

$$\Delta\theta = \arccos\left(\frac{B_X^2 + B_Y^2 + 1}{\sqrt{(B_X + 1)^2 + B_Y^2}\sqrt{(B_X - 1)^2 + B_Y^2}}\right) .$$

Figure 5(b) shows the phase shift for the first quadrant of the $B_X$-$B_Y$-plane. This plot has mirror symmetry for both coordinate axes, though, for a mirroring relative to the $B_X$-axis the phase shift would become negative. As the phase shift refers to the angle between **P** and **N** its value is always positive and $\leq 180°$ per definition.

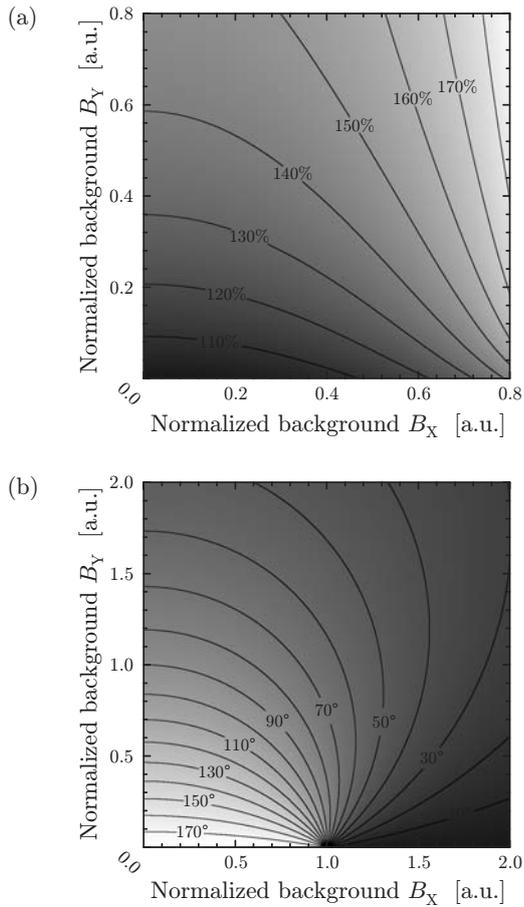

**Fig. 5.** (a) Simulation of the full width at half maximum $W$ of an antiparallel domain wall imaged with the $R$-output while a background **B** is present. The width is normalized to its original value. (b) Calculated phase shift $\Delta\theta$ at an antiparallel domain wall for a mixed background with components $B_X$ and $B_Y$. Both plots have mirror symmetry at the $B_X$- and the $B_Y$-axis why the first quadrant is shown only.

**Background-free PFM imaging**

Now the crucial point is: how can one get reliable data from PFM imaging despite the background signal? Firstly, one should be aware that getting rid of the background seems difficult. Different mountings and thus the mechanical coupling of the cantilevers to the PFM head alter but not suppress the background. The elongation of the tube scanner also influences the background; that is why PFM scans on tilted surfaces sometimes got a linear offset. Insufficient electrical shielding of the PFM head probably has an influence too. However, the origin of the PFM background is still unknown. Therefore it is not clear how to suppress it. Fortunately, there is no need for suppressing the background to record reliable experimental data in PFM imaging because straight information can be obtained when using the $X$-output of the lock-in amplifier, as has been demonstrated in this work.

From the experimental side, there is an additional problem arising: Theoretically, the vibration of the surface due to the converse piezoelectric effect should be in-phase with the alternating voltage applied to the tip (at least for frequencies < 100 kHz). Therefore the piezoresponse signal **d** should only show up in the $X$-output of the lock-in amplifier. There is, however, always also a small contribution in the $Y$-signal, probably because of an electronically governed phase shift of the SFM, which has been observed by others too (Eng et al., 1998). In the vector diagram of Fig. 1(a) the piezoresponse signal **d** would show up slightly tilted. To extract nevertheless correct data from PFM measurements the easiest solution is to set the phase of the lock-in amplifier such that no domain contrast is visible in the $Y$-output (thus again **d** is parallel to the $X$-axis). This corresponds to a rotation of the coordinate system in the vector diagram of Fig. 1(a). An equivalent (and even more precise) solution is a rotation of the coordinate system after image acquisition such that the standard deviation of the $Y$-image is minimized.

**Conclusions**

In conclusion we have analyzed the influence of the system inherent background on the images of ferroelectric domains obtained with piezoresponse force microscopy concerning domain contrast and domain boundaries. We have pointed out possible origins for misinterpretation of PFM images when using the magnitude output of the lock-in amplifier for readout. Finally, we recommended a detection scheme to get reliable data in PFM imaging.

**Acknowledgements**

The authors thank G. Baldenberger from the Institut National d'Optique (Canada) for providing the PPLN samples. Financial support of the DFG research unit 557 and of the Deutsche Telekom AG is gratefully acknowledged.

**References**

Alexe, M. & Gruverman, A. (2004) *Nanoscale Characterisation of Ferroelectric Materials.* Springer, Berlin, New York.

Agronin, A., Molotskii, M., Rosenwaks, Y., Strassburg, E., Boag, A., Mutchnik, S., & Rosenman, G. (2005) Nanoscale piezoelectric coefficient measurements in ionic conducting ferroelectrics. *J. Appl. Phys.* **97**, 084312.

Barry, I.E., Ross, G.W., Smith, P.G.R. & Eason, R.W. (1999) Ridge waveguides in lithium niobate fabricated by differential etching following spatially selective domain inversion. *Appl. Phys. Lett.* **74**, 1487–1488.

Broderick, N.G.R., Ross, G.W., Offerhaus, H.L., Richardson, D.J., & Hanna, D.C. (2000) Hexagonally poled lithium niobate: A two-dimensional nonlinear photonic crystal. *Phys. Rev. Lett.* **84**, 4345–4348.

Cho, Y., Hashimoto, S., Odagawa, N., Tanaka, K. & Hiranaga, Y. (2005) Realization of 10 Tbit/in$^2$ memory density and subnanosecond domain switching time in ferroelectric data storage. *Appl. Phys. Lett.* **87**, 232907.

Eng, L.M., Güntherodt, H.J., Rosenman, G., Skliar, A., Oron M., Katz, M. & Eger, D. (1998) Nondestructive imaging and characterization of ferroelectric domains in periodically poled crystals. *J. Appl. Phys.* **83**, 5973–5977.

Fejer, M.M., Magel, G.A., Jundt, D.H. & Byer, R.L. (1992) Quasi-phase-matched second harmonic generation: Tuning and tolerances. *IEEE J. Quantum Elect.* **28**, 2631–2654.

Gahagan, K.T., Scrymgeour, D.A., Casson J.L., Gopalan, V., & Robinson, J.M. (2001) Integrated high power electro-optic lens and large-angle deflector. *Appl. Opt.* **40**, 5638–5642.

Harnagea, C., Alexe, M., Hesse, D. & Pignolet, A. (2003) Contact resonances in voltage-modulated force microscopy. *Appl. Phys. Lett.* **83**, 338–240.

Hong, S., Shin, H., Woo, J. & No, K. (2002) Effect of cantilever-sample interaction on piezoelectric force microscopy. *Appl. Phys. Lett.* **80**, 1453-1455.

Jazbinšek, M. & Zgonic, M. (2002) Material tensor parameters of LiNbO$_3$ relevant for electro- and elasto-optics. *Appl. Phys. B* **74**, 407–414.

Jungk, T., Hoffmann, A. & Soergel, E. (2006) Quantitative analysis of ferroelectric domain imaging with piezoresponse force microscopy. *Appl. Phys. Lett.* **89**, 163507.

Jungk, T., Hoffmann, Á. & Soergel, E. (2007) Impact of the tip radius on the lateral resolution in piezoresponse force microscopy. *arXiv:cond-mat* 0703793.

Kolosov, O., Gruverman, A., Hatano, J., Takahashi, K. & Tokumoto, H. (1995) Nanoscale visualiza-tion and control of ferroelectric domains by atomic force microscopy. *Phys. Rev. Lett.* **74**, 4309–4312.

Labardi, M., Likodimos, V. & Allegrini, M. (2000) Force-microscopy contrast mechanisms in ferro-electric domain imaging. *Phys. Rev. B* **61**, 14390–14398.

Labardi, M., Likodimos, V. & Allegrini, M. (2001) Resonance modes of voltage-modulated scanning force microscopy. *Appl. Phys. A* **72**, S79–S85.

Landolt Börnstein (1981) *Numerical data and functional relationships in science and technology* Group **III** Volume **16**, Springer, Berlin; New York.

Rabe, U., Janser, K. & Arnold, W. (1996) Vibrations of free and surface-coupled atomic force microscope cantilevers: Theory and experiment. *Rev. Sci. Instrum.* **67**, 3281–3293.

Scrymgeour, D.A. & Gopalan, V. (2005) Nanoscale piezoelectric response across a single antiparallel ferroelectric domain wall. *Phys. Rev. B* **72**, 024103.

Shvebelman, M., Urenski, P., Shikler, R., Rosenman, G, Rosenwaks, Y, & Molotskii, M. (2002) Scanning probe microscopy of well-defined periodically poled ferroelectric domain structure. *Appl. Phys. Lett.* **80**, 1806–1808.

Soergel, E. (2005) Visualization of ferroelectric domains in bulk single crystals. *Appl. Phys. B* **81**, 729–752.

Xu, C.H., Woo, C.H., Shi, S.Q. & Wang, Y. (2004) Effects of frequencies of AC modulation voltage on piezoelectric-induced images using atomic force microscopy. *Mater. Charact.* **52**, 319-322.

Zhang, X., Hashimoto, T. & Joy, D.C. (1992) Electron holographic study of ferroelectric domain walls. *Appl. Phys. Lett.* **60**, 784–786.